# Self-aligned process for forming microlenses at the tips of vertical silicon nanowires by atomic layer deposition

Yaping Dan[1*], Kaixiang Chen[1], and Kenneth B. Crozier[2]

[1]University of Michigan – Shanghai Jiao Tong University Joint Institute, Shanghai Jiao Tong University, Shanghai, China 200240

[2]School of Engineering and Applied Sciences, Harvard University, Cambridge, MA, USA, 02138

*To whom the correspondence should be addressed: yaping.dan@sjtu.edu.cn

*Abstract*

*The microlens is a key enabling technology in optoelectronics, permitting light to be efficiently coupled to and from devices such as image sensors and light-emitting diodes. Their ubiquitous nature motivates the development of new fabrication techniques, since existing methods face challenges as microlenses are scaled to smaller dimensions. Here, we demonstrate the formation of microlenses at the tips of vertically-oriented silicon nanowires via a rapid atomic layer deposition (ALD) process. The nature of the process is such that the microlenses are centered on the nanowires, and there is a self-limiting effect on the final sizes of the microlenses arising from the nanowire spacing. Finite difference time domain electromagnetic simulations are performed of microlens focusing properties, including showing their ability to enhance visible-wavelength absorption in silicon nanostructures.*

Microlenses are important components for micro/nano scale optics and optoelectronics (*1-10*). Arrays of microlenses are used in image sensors to focus light onto the photodetectors (*1*), and in photonic integrated circuits (IC) to collimate light from light emitting diodes (LED)(*7, 10*) and lasers(*4*). Interestingly, Lee and colleagues(*11*) recently discovered that near-field focusing by microlenses can beat the diffraction limit. Today's commercial technology for microlens fabrication relies on the reflow of patterned photoresist films at elevated temperature(*12*). This method, however, become more difficult as the microlenses are scaled to smaller sizes. First, there is minimum gap size that must be maintained between the lenses to ensure high device yield. If this gap size is not maintained, i.e. the lenses come physically into contact, the lenses will merge during the thermal reflow process and become defective. The existence of this gap means that some of the light impinging upon the image sensor is not focused onto the photodetectors. This effect becomes increasingly problematic as pixels (and microlenses) are scaled to smaller dimensions. Second, in the photoresist reflow method, the alignment between the microlenses and the corresponding optoelectronic devices must be very carefully controlled. The implications of misalignment become of course more serious as devices are scaled down. To address these issues, we here demonstrate a new microlens fabrication technology based on a rapid $SiO_2$ atomic layer deposition (ALD) process. An array of microlenses is formed on an array of vertically-oriented nanowires by the conformal coating property of ALD. The self-alignment nature of this method ensures that the microlenses are centered on the nanowires. The size and shape of the microlenses depend on the spacing between the nanowires and the thickness of the deposited films. This method presents the opportunity for microlenses to be further scaled down and used in next generation optoelectronic devices and systems.

Our microlenses are formed by conformal coating of $SiO_2$ around vertically-oriented silicon nanowires (Fig.1a) that are fabricated through inductively coupled plasma-reactive ion etching (ICP-RIE). This top-down approach for nanowire fabrication is based on the Bosch process(*13, 14*), but with modifications. The Bosch method is a cyclic process containing etching and passivation phases. In the etching phase, $SF_6$ gaseous molecules are ionized by the radiofrequency electric field to form reactive $SF_5^+$ plasma that isotropically etches silicon. Etching in the vertical direction comes along with an undercut below the etch mask as shown in Fig.1b(i). To mitigate the undercut, the process is switched to the passivation phase by cutting off $SF_6$ and turning on $C_4F_8$. The ionized $C_4F_8$ molecules form a polymeric coating all over the surfaces (Fig.1b(ii)), preventing further etching of the sidewalls as well as the bottom surface. To allow the silicon at the bottom to be selectively etched, a DC bias is applied in the vertical direction to accelerate the ions towards the substrate, which results in physical bombardment of the polymer on the bottom surface. This bombardment exposes the silicon at the bottom by removing the polymer coating (Fig.1b(iii)), while silicon on the side wall remains protected by the polymer. After the process switches back to the etching phase in Fig.1b(i), isotropic etching again occurs. If repeated over multiple cycles, the process permits the fabrication of vertically-oriented nanowires with curvy sidewalls (*13, 15*). To obtain smoother surfaces, we modify the Bosch process by combining the etching and passivation phases into a single phase, thereby allowing the silicon to be etched and passivated at the same time. The two competing phases can be tuned by adjusting the flow rate ratio of $SF_6$ and $C_4F_8$. By choosing this ratio appropriately(*16, 17*), vertically-oriented nanowires with smooth surface morphology can be achieved (Fig.1a).

To form the microlenses on tops of the nanowires, we place the nanowire sample in an ALD chamber for conformal $SiO_2$ coating. ALD is often regarded as a slow process, a consequence of the fact that the material is grown layer by layer. One might therefore expect microlenses to take days to fabricate by ALD. Here, however, we employ a rapid ALD process to grow $SiO_2$ in which deposition rates can be up to ~12 nm per cycle(*18*). Using this method, we demonstrate fabrication of arrays of microlenses (with diameters ~2 μm) in several hours. The ALD method uses liquid tris(tert-pentoxy)silanol (TPS) as the precursor that is catalyzed by gaseous trimethyl aluminum (TMA). In the process, TMA first reacts with available hydroxylated sites (Fig.2a) on the silicon surface, followed by the reaction of silanols at the Al center (Fig.2b). Additional silanol precursors will further insert at the Al catalytic center (Fig.2c) by diffusing to the bottom. At the same time, the siloxane chains cross-link to form $SiO_2$ (Fig.2d). The competition between the diffusion and the cross-linking processes results in the $SiO_2$ deposited in each cycle being limited to a certain thickness.

We employ the rapid ALD process to fabricate the microlenses by conformally coating a 1μm thick $SiO_2$ layer on vertically-oriented nanowires. To demonstrate the influence of nanowire pitch on the final lens morphology, nanowire arrays with a variety of pitches are used. Fig.3a, b and c show nanowire arrays with pitches of 2μm, 1.5 μm and 1 μm, respectively. Highly uniform microlenses arrays are formed. For the nanowire array with 2μm pitch, the microlenses close to 2μm in diameter are almost in contact. As indicated by the dashed circle and inset, the array shown in Fig.3a contains an additional nanowire located between two normally-positioned nanowires. It can be seen that the microlens that results has the shape of a dumbbell. The underlying principle will be seen more clearly in the next few figures. For the nanowires with 1.5μm picth (Fig.3b), the lenses have started to overlap, with gaps formed between the nanowires. Each gap is centered at the intersection between diagonal lines connecting the nanowires, as shown in the inset of Fig.3b. For those with 1 μm pitch, the lenses have completely merged, resulting in each lens occupying an area of 1μm × 1μm (Fig.3c). Clearly, the final size and shape of each lens is dependent on the positions of the underlying nanowires. The fabrication process is schematically depicted in Fig.3d.

To further investigate the process that forms the microlenses, we employ a focused ion beam (FIB) to cut the lenses in half, thereby exposing their cross-sections (Fig.4a). It can be clearly seen that each microlens has a vertical nanowire in its center, and that $SiO_2$ conformally coats the nanowire in accordance with the schematic illustration of Fig.3d. We perform electromagnetic simulations (finite difference time domain method, FDTD) to study the focusing properties of the microlenses. For simplicity, two-dimensional simulations of the absorption cross sections of silicon ridges (rather than three-dimensional simulations of nanowires) are performed. It can be seen that the microlens focuses collimated illumination to a spot located approximately 2μm away from the lens top (Fig.4b). We next study the consequences of the focusing provided by the microlenses upon the nanowire absorption. Simulations are performed for nanowires embedded in (cylindrical) lenses, and for nanowires embedded in thin films with flat surfaces. The results are shown as Fig.4c. At shorter wavelengths, the absorption is enhanced more than 3 times by the microlens. The peak at the wavelength of 430 nm is due to the excitation of a waveguide mode of the ridge, and the resultant light absorption[16, 19]. There has recently been interest[20, 21] in the use of vertically-oriented silicon nanowires for imaging, including those with integrated PN junctions for photocurrent collection(*22*). Ideally, for such devices one would like to increase the absorption by the nanowire, and the results of Fig. 4c show that the integrated microlenses present a means for doing this. In the reciprocal process, the microlenses could be similarly used to collimate light emitted by nanowire devices. While nanowire-based LEDs and lasers have been investigated intensively in recent years, few investigations have been made of techniques to collimate their output. Fig. 4d shows an optical microscope image of the microlens array containing embedded nanowires. From the zoom-in image (inset of Fig. 4d), it can be seen that a dark spot, with a blue-ish tint, appears at the center of each microlens. Each spot is believed to correspond to a nanowire, as it is the only object in this location. Interestingly, it should be noted that the presence of the microlens is expected to alter the magnification of the image of each nanowire. For example, when a hemisphere lens (solid immersion lens) is added to an optical microscope, the transverse magnification of the image is increased $n$ times, where $n$ is the lens refractive index (*23*). Note that this refers to the case where the object is placed on the bottom flat surface of the hemisphere.

In conclusion, we demonstrated a microlens fabrication technology using a rapid ALD process. By conformally coating $SiO_2$ around the vertically-oriented nanowires, microlenses are produced on the tops of the nanowires. The microlenses have shapes and dimensions that are determined by the spatial distribution of the nanowires. This technology offers a promising solution for very small microlenses that may find a wide range of applications in next generation advanced optoelectronic systems.

Acknowledgement

The work is supported by the "Pujiang Talent Program" of Shanghai Municipal Government (GJ3700001), National Science Foundation of China (BC3700019), and "1000 Young Scholar Plan" of Chinese Central Government (GKKQ3700001). This work was also supported by Zena Technologies and the National Science Foundation of the United States of America (NSF, grant ECCS-1307561). The experiments were conducted at the Center for Nanoscale Systems (CNS) at Harvard University, which is supported by the NSF.

Captions:

Figure 1. Fabrication process for vertically-oriented nanowires. (a) Scanning electron microscope (SEM) image of vertically-oriented silicon nanowires. Nanowires are typically ~100-300 nm in diameter and several micrometers long. (b) Bosch process cycle: (i) Isotropic etching of silicon by ionized $SF_6$ gaseous molecules; (ii) polymeric coating by $C_4F_8$; (iii) Polymeric coating at the bottom surface is removed by physical bombardment of ions. (c) Combination of etching and passivation improves surface morphology of nanowires.

Figure 2. ALD process for rapid deposition of $SiO_2$. (a) TMA reacts with available hydroxylated sites on substrate. (b) Silanols react at Al center. (c) Additional silanol precursors further insert at Al catalytic centers by diffusing to bottom. (d) Cross-linking of siloxane chains prevents silanol precursors from diffusing to bottom and therefore stops growth of $SiO_2$ until a new cycle begins. This figure is modified from a figure of Ref.(*18*).

Figure 3. Top view of SEM images of microlens arrays. (a) Lenses are isolated from one another. Inset: defect in nanowire array modifies shape of microlens. (b) Lenses partially overlap. Inset: diagram showing how gaps form between lenses. (c) Lenses have merged. (d) Diagram showing conformal coating of $SiO_2$ around nanowires. Final sizes and shapes of lenses can be controlled by appropriate choice of nanowire position.

Figure 4 (a) SEM (tilted view) of microlens cross-section after it has been cut by FIB. Inset: diagram showing how SEM image is obtained at a tilted angle. (b) Periodic two dimensional FDTD simulation of electric field intensity ($|E|^2$) on cross-section of cylindrical $SiO_2$ microlens resulting from normally-incident TE plane wave illumination from the top ($\lambda$= 700 nm). (c) Periodic two dimensional FDTD simulations of absorption cross-sections of silicon nanoridge encased in cylindrical $SiO_2$ microlens, and silicon nanoridge encased in $SiO_2$ with flat top surface. Nanoridge is 100 nm wide and 2μm tall. Radius of curvature of microlens is 1μm. (d) Optical microscope image of microlens array. Close-up inset shows one blue-ish dark spot, believed to be the underlying nanowire in each microlens.

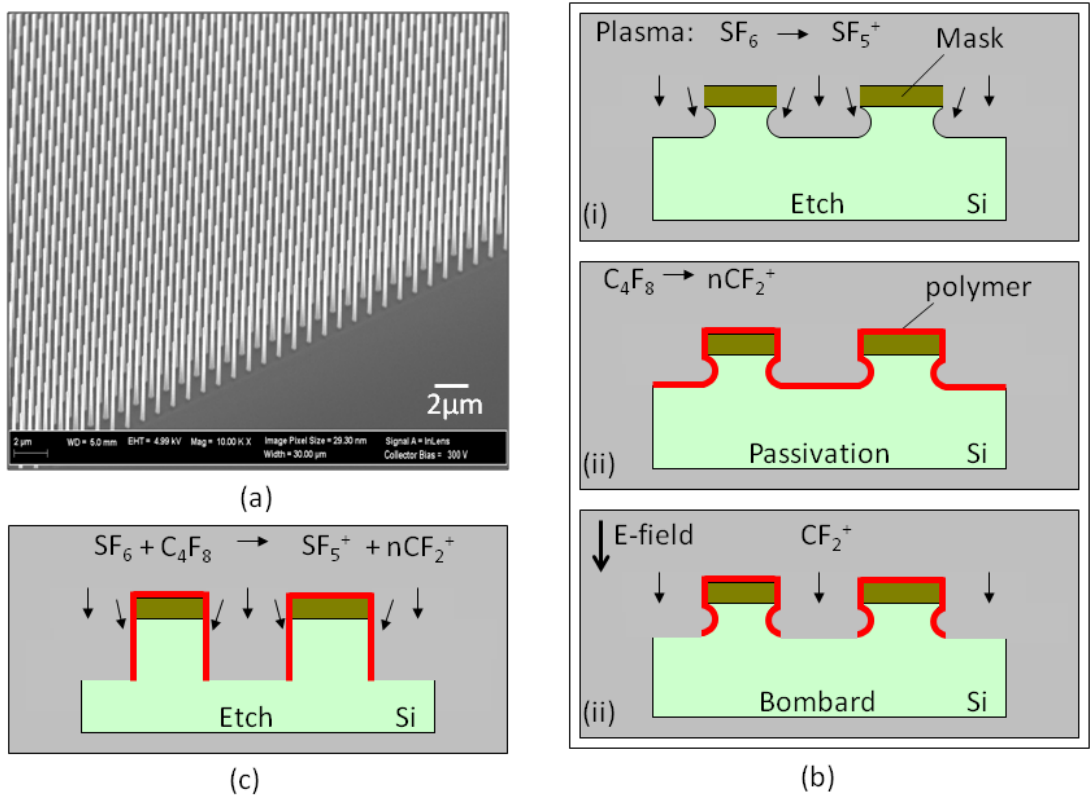

Figure 1.

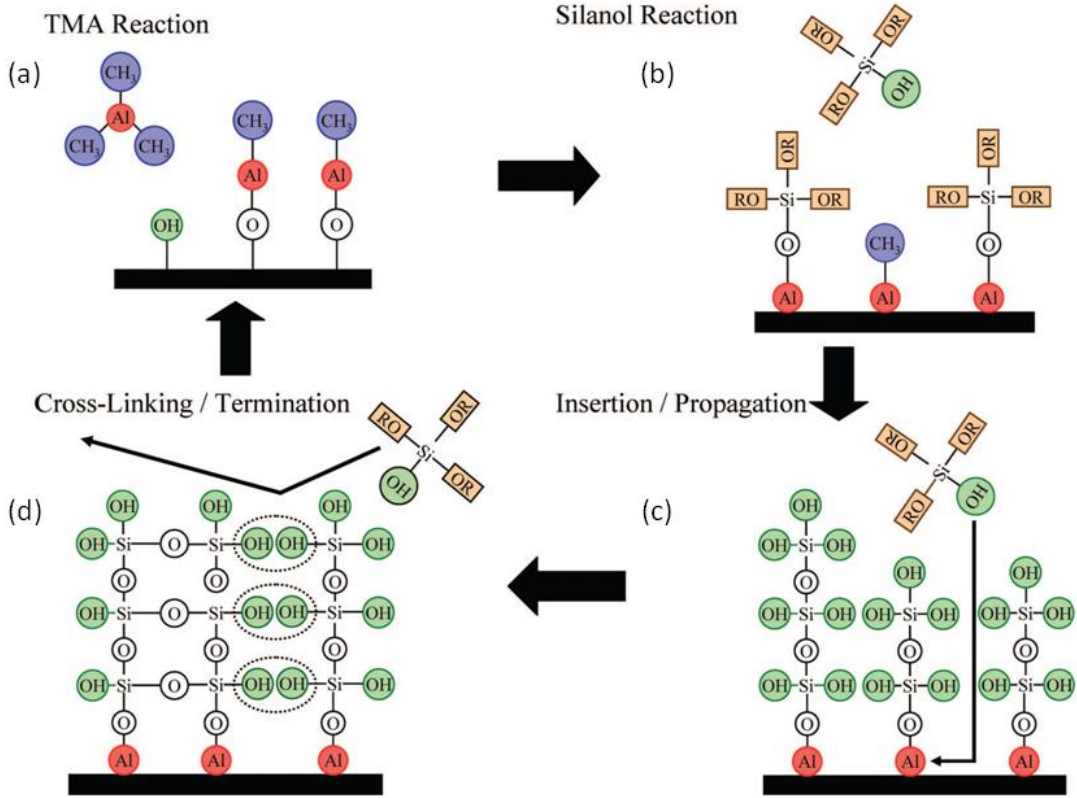

Figure 2.

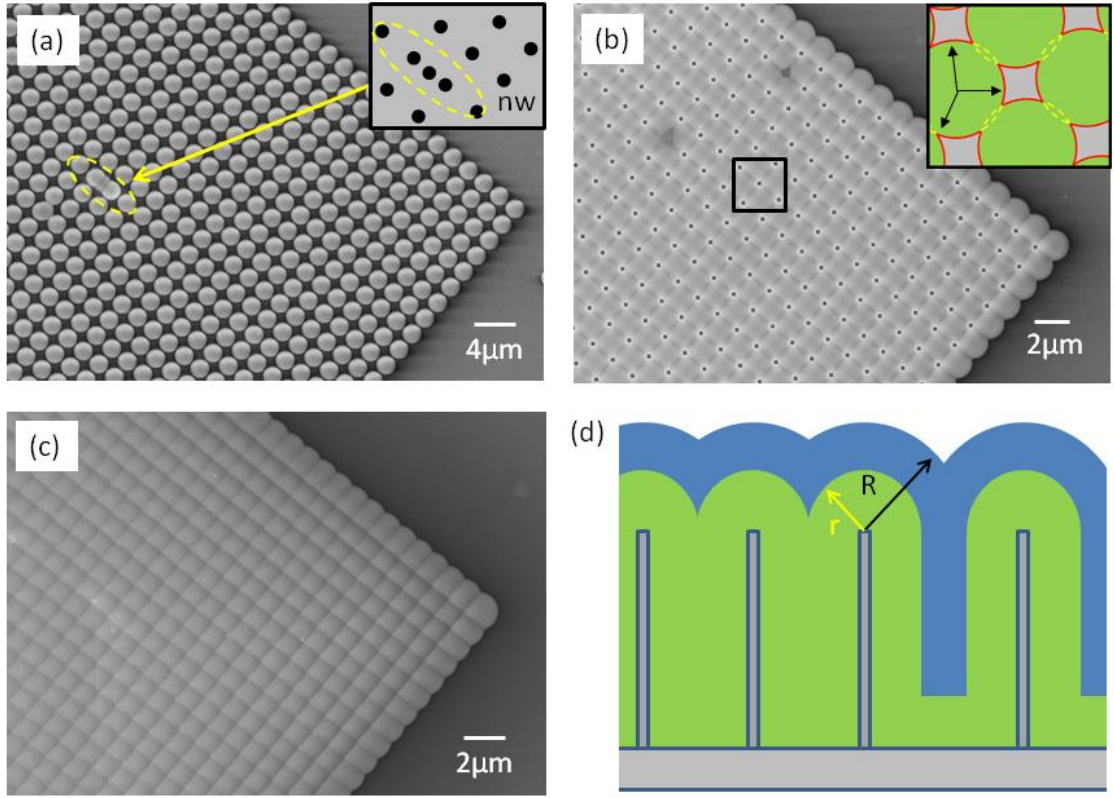

Figure 3.

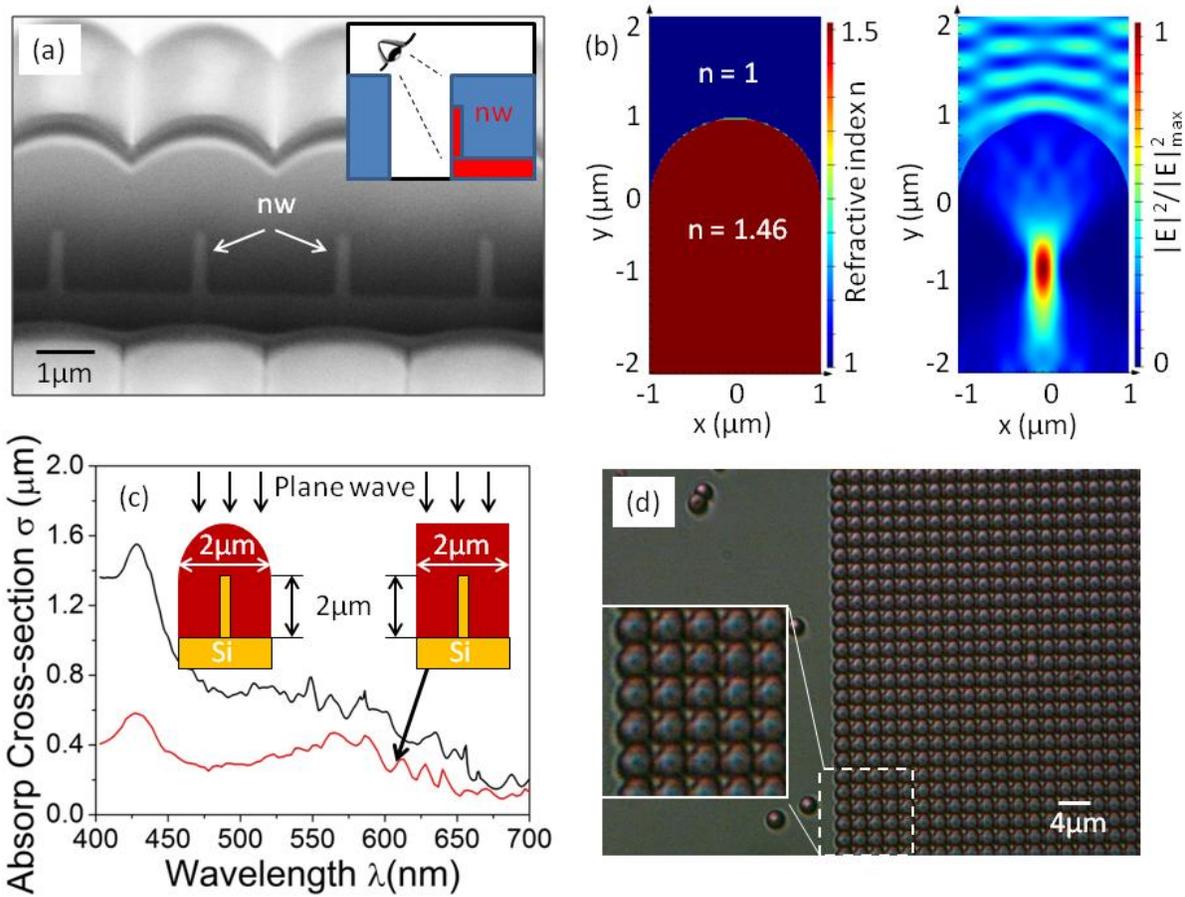

Figure 4